# Star-Mesh Quantized Hall Array Resistance Devices

Dean G. Jarrett*[1], Ching-Chen Yeh[1,2], Shamith U. Payagala[1], Alireza R. Panna[1], Yanfei Yang[3], Linli Meng[3], Swapnil M. Mhatre[1,2], Ngoc Thanh Mai Tran[1,4], Heather M. Hill[1], Dipanjan Saha[1], Randolph E. Elmquist[1], David B. Newell[1], and Albert F. Rigosi[1]

[1]*Physical Measurement Laboratory, National Institute of Standards and Technology (NIST), Gaithersburg, MD 20899, United States*

[2]*National Taiwan University, Taipei 10617, Taiwan*

[3]*Graphene Waves, LLC, Gaithersburg, MD 20899, United States*

[4]*Joint Quantum Institute, University of Maryland, College Park, MD 20742, United States*

ABSTRACT: Advances in the development of graphene-based technology have enabled improvements in DC resistance metrology. Devices made from epitaxially grown graphene have replaced the GaAs-based counterparts, leading to an easier and more accessible realization of the ohm. By optimizing the scale of the growth, it has become possible to fabricate quantized Hall array resistance standards (QHARS) with nominal values between 1 kΩ and 1.29 MΩ. One of these QHARS device designs accommodates a value of about 1.01 MΩ, which made it an ideal candidate to pursue a proof-of-concept that graphene-based QHARS devices are suitable for forming wye-delta resistance networks. In this work, the 1.01 MΩ array output nearly 20.6 MΩ due to the wye-delta transformation, which itself is a special case of star-mesh transformations. These mathematical equivalence principles allow one to extend the QHR to the 100 MΩ and 10 GΩ resistance levels with fewer array elements than would be necessary for a single array with many more elements in series. The 1.01 MΩ device shows promise that the wye-delta transformation can shorten the calibration chain, and, more importantly, provide a chain with a more direct line to the quantum SI.

# I. INTRODUCTION

Quantized Hall array resistance standards (QHARS) are devices that have been designed to accommodate many smaller elements that each output a resistance that is a multiple, integer or fractional, of $h/e^2$, where $h$ is the Planck constant and $e$ is the elementary charge, respectively. Historically, QHARS were made with GaAs/AlGaAs heterostructures until they were replaced by graphene for metrology applications in the United States [1] – [3]. The ease of use associated with using graphene-based devices quickly caught on [4] – [6], with many groups worldwide using epitaxially grown graphene (EG) as a quantized Hall resistance standard. Most graphene-based standards operate at the resistance plateau formed by the $v = 2$ Landau level (about 12906.4037 Ω) since that plateau is easier to access than the $v = 6$ plateau or others exhibited by graphene [7].

Assembling series and parallel connections of many Hall bars is now a promising avenue of research due to improved device geometries and superconducting electrical contacts between elements [8] – [10]. By adding sufficient elements in series, one may be able to shorten the chain of calibration by having higher quantized resistances. For instance, an array device valued at around 1 MΩ would require a minimum of 78 elements. Though feasible, engineering issues for even higher resistances compound rapidly since those higher decades, namely those 10 MΩ and beyond, would require an order of magnitude increase in the number of elements. For instance, it would require approximately 7748 array elements in series (assuming $v = 2$ quantization) to make an array nearly 100 MΩ. This rapidly growing number of required elements for higher resistances presents a formidable engineering challenge.

To circumvent this scaling problem, QHARS devices were constructed with designs suitable for use in a wye-delta (Y-Δ) network. QHARS have been used for several efforts in resistance metrology, both of the graphene and GaAs/AlGaAs variety [8] – [14]. The exemplary 1.01 MΩ device has two arrays of 39 elements each connected in series with a single element connected at the midpoint to provide a way to check quantization of each 39-element array. These three arms of the 1.01 MΩ device, by using the Y-Δ transformation, form higher resistance standards when compared to the three, relatively smaller, components. Due to the electrical and mathematical equivalence of the wye and delta networks, this transformation can be used to construct standards with values between megaohms and gigaohms [15] – [16]. The idea of using QHARS to form a Y-Δ transfer standard may be expanded to include future QHARS devices with transformed values of 100 MΩ and 10 GΩ with only several hundred elements, far fewer than the much larger numbers of 7748 to $7.75 \times 10^5$ devices in series, respectively.

For this work, several 1.01 MΩ devices were fabricated at the National Institute of Standards and Technology (NIST) for calibrating 10 MΩ, 100 MΩ, and 1 GΩ resistance standards directly with a two-terminal cryogenic current comparator (CCC) [17] in a single step, without having to do two steps from a single Hall bar element. A dual source bridge (DSB) is also employed to measure the equivalent Y-Δ resistance of about 20.6 MΩ, proving that this overall concept is beneficial to future resistance metrology applications.

## II. Device fabrication and characterization

Graphene films were grown on 22.8 mm × 22.8 mm silicon carbide chips. The chip was diced from a semi-insulating SiC wafer of diameter 10.2 cm (about 4 in) from Wolfspeed (see Acknowledgments for commercial disclaimer and Appendix for growth information). The sample was cleaned with Piranha solution (3:1 $H_2SO_4$:$H_2O_2$) for 33 minutes at 120 °C, followed by a 5 min clean with 51 % hydrofluoric acid (by volume and diluted with deionized water). Moments before the growth process was initiated, the chip was coated with a dilute solution of carbon-based photoresist (AZ 5214E, see Acknowledgments) in isopropanol to take advantage of the benefits of polymer-assisted sublimation growth (PASG) [18]. The graphite-lined resistive-element furnace (Materials Research Furnaces Inc., see Acknowledgments) was flushed with Ar gas and filled to about 103 kPa from a 99.999 % liquid argon source before being held at about 1850 °C for 4 min [19] – [20]. The chip with grown EG was removed after the system was allowed to cool to room temperature.

The grown EG samples were characterized using both optical and confocal laser scanning microscopy (CLSM). High-resolution confocal images have been taken at more than 10 sampling sites (marked by the orange, green and red squares in Fig. A1 in the Appendix) for a quick evaluation of the variation of graphene thickness across the chip. More coverage information is available in the Appendix. Eight 1.01 MΩ devices have been fabricated in the region with minimum multilayers.

The device fabrication is similar to others reported in recent papers, whereby the EG layer has a 20 nm layer of Pd/Au deposited on it, followed by photolithography processes for defining the Hall bar and device contacts [7], [21] – [22]. Though the intrinsic electron density in epitaxial graphene on SiC is near $10^{13}$ cm$^{-2}$, it is greatly reduced after the Pd/Au layer is removed by aqua regia [7], due to a *p*-doping process by the nitric acid [22]. The 1.01 MΩ devices are exposed to ambient air after fabrication so that the adsorption of oxygen molecules from the air will further *p*-dope graphene below $10^{11}$ cm$^{-2}$. Gently annealing the devices in vacuum at a temperature of about 85 °C will release oxygen molecules slowly and the desired carrier density can be obtained by controlling the annealing time [22].

For the electrical contacts of the QHARS devices, a layer of superconducting NbTiN was deposited to greatly improve array performance [9]. Moreover, the contacts' design of incorporating a multi-series connection was critical to device functionality (see Appendix for optical and CLSM images of the device), namely, to eliminate uncertainty due to lead resistances and to optimize the current flow [9]. The separation of the NbTiN layer and the EG was greater than 80 nm so that undesired quantum effects, such as Andreev reflection, could be prevented.

Testing material homogeneity is crucial for ensuring a fully quantized device. After the first inspection done during the fabrication process, which involved CLSM and optical microscopy, a second, noninvasive inspection for homogeneity was performed via Raman spectroscopy given the potentially high doping [23] – [24]. Optical properties of the EG also give an insight into the quality of the material that could have been overlooked. Raman measurements were performed with a Renishaw InVia micro-Raman spectrometer (see Acknowledgements). A helium-neon laser, with excitation wavelength of 633 nm, was used as the source. Each spectrum was measured using a backscattering configuration, 2 μm spot size, 1 mW power, 50 × objective, 300 s acquisition time, and 1200 mm$^{-1}$ grating. More information is provided in the Appendix.

### III. Verification and Measurement methodology

#### A. Intended Device Functionality

The aforementioned 1.01 MΩ device is intended to act as an unknown resistor, or rather, a resistor whose value is to be determined through this experiment and compared with what its quantized value should be. Each of its two arrays, composed of 39 elements each and connected in series, meet at a common node with a single element. The two equivalent arms nearly 0.5 MΩ each, along with the single element, make up the three resistors of a Y-network (designated $R_X$, Hi–Lo–Gnd, or $R_1$–$R_2$–$R_0$) and can be equated to three resistors arranged as a triangular mesh containing one less node than the Y-network (designated $R$, $R_a$, and $R_b$, where the latter two are inconsequential to the desired measurement). This equivalence is shown in Fig. 1 (a) and is the essence of the Y-Δ transformation [15], which itself is a special case of star-mesh transforms [25] – [27]. These star-mesh transforms are used to reduce the number of nodes by one. More details on cases beyond the triangular (Δ) cases will be provided later. Depending on the values of the Y resistors, one can achieve higher, equivalent, quantized resistances with simple mathematical formulae:

$$R = \frac{R_1 \times R_2}{R_0} + R_1 + R_2$$

$$R_a = \frac{R_1 \times R_0}{R_2} + R_1 + R_0$$

$$R_b = \frac{R_2 \times R_0}{R_1} + R_2 + R_0$$

(1)

It follows from Eq. 1 that in order to maximize the transformation for $R$, it would benefit greatly if one were to minimize $R_0$ given that it is the denominator. The $R_0$ arm is thus, arguably, the most influential piece of the Y configuration.

When configured properly and calculated via Y-Δ transform as shown in Fig. 1 (b), the QHARS device yields an equivalent resistance $R$ of about 20.6 MΩ. This higher value could be used for calibrating 100 MΩ and 1 GΩ high resistance standards with a DSB having a 5:1 or 50:1 ratio, respectively [15], [28]. The 1.01 MΩ device is also designed to eventually be used with a two-terminal CCC as a means of scaling directly to higher resistances of 10 MΩ, 100 MΩ, and 1 GΩ with CCC turn-winding ratios of 10:1 or 100:1 [17], [29].

With relatively minor modifications in fabrication, similar array networks can be made for resistance values closer to decade values. Table 1 shows a few possible Y-Δ transformations that could generate a resistance $R$ for a QHARS device of corresponding design elements. These future QHARS devices may yield values closer to 100 MΩ and up to 10 GΩ using the Y-Δ transformation. In the case of a 10 GΩ equivalent resistance, a QHARS device would need to accommodate 501 elements, which is not an unreasonable projection given recent developments using several hundred [14]. The $R_1$ and $R_2$ arms in Table 1 mainly have elements in series, but it is possible to introduce one or more smaller parallel resistors (increasing the number of nodes – see Appendix) to finely tune the desired equivalent resistance.

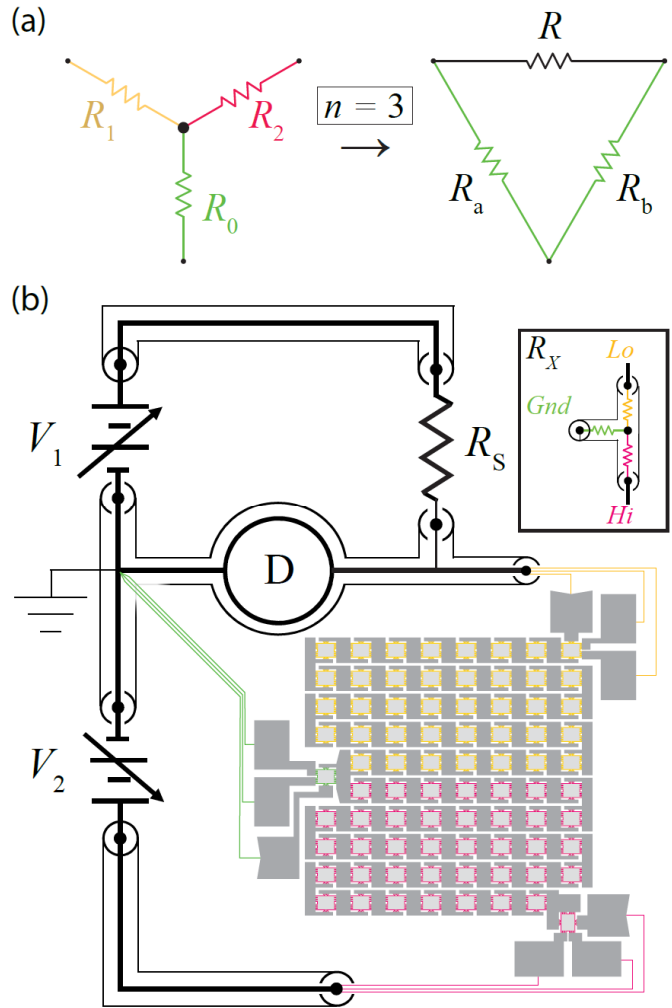

Fig. 1. (a) Illustration of the Y-Δ transformation is provided, reflecting the experimental setup. This transformation is a special case of the star-mesh transformation, which reduce the number of nodes by generating an equivalence resistance network. (b) Simplified diagram for experimental methods involving the use of a dual source bridge (DSB). The top arm applies a voltage $V_1$ across a known reference resistance $R_S$, while the lower arm applies voltage $V_2$ with opposite polarity across an unknown resistor $R_X$. The voltage is then modified until the detector (labelled D) reads a null signal. The 1.01 MΩ device is intended to substitute $R_X$ (see upper inset), with each of its two arrays, composed of 39 elements each and connected in series, meeting at a common node with a single element. The single element represents the *Gnd* (or $R_0$, color coded green), whereas the two larger arms make up the *Lo* and *Hi* ($R_1$ and $R_2$, respectively) terminal connections. It should be noted that a typical DSB setup has the unknown resistance $R_X$ on the top circuit and the $R_S$ on the bottom circuit.

TABLE I
Y-Δ TRANSFORMATIONS FOR FUTURE QHARS DEVICES

| $R_1$ (elements) | $R_2$ (elements) | $R_0$ (elements) | Total (elements) | R (MΩ) |
|---|---|---|---|---|
| 39 | 39 | 2 | 80 | 10.8220 |
| 39 | 39 | 1 | 79 | 20.6373 |
| 50 | 50 | 1 | 101 | 33.5566 |
| 60 | 60 | 1 | 121 | 48.0118 |
| 80 | 80 | 1 | 161 | 84.6660 |
| 96 | 79 | 1 | 176 | 100.141 |

### B. The Dual Source Bridge and Transport Setup

To validate predictions obtained with the Y-Δ transformations, most measurements were performed using a DSB. Figure 1 (b) shows a DSB, also known as a modified Wheatstone bridge, which has been implemented in the past at various National Metrology Institutes [28] – [31]. Generally, on the top arm, a voltage $V_1$ may be applied across a known reference resistance $R_S$, while on the lower arm, a voltage $V_2$ may be applied with opposite polarity across an unknown resistor $R_X$. In a DSB, $R_X$ or $R_S$ may be in either the upper or lower arm since $V_1$ and $V_2$ are interchangeable programmable voltage sources. Here the reference resistor $R_S$ is used to evaluate the QHARS $R_X$. To calibrate a higher value standard resistor, the QHARS would be the standard $R_S$ and a high value resistor would be the unknown $R_X$. The voltage is then adjusted until the detector (labelled D) reads a null signal.

A significant benefit from using a DSB is the very low uncertainties that can be achieved due to the simple calibration of the applied voltages. Additionally, leakage effects become negligible since the sensitive bridge point detector gets balanced to a null current and the low impedance (< 0.1 Ω at DC) of the voltage sources. The main uncertainties are the calibration of the voltage sources, offset voltages, noise, and the reference resistor $R_S$. Accurate measurements of Y-networks (also called T-networks) using a DSB require the *Lo* terminal $R_2$ to be at the same potential as that of the *Gnd* terminal on $R_0$. Tetrahedral junctions [32] have been used to connect the three sets of triple-series leads from the QHARS to the DSB. By adding another tetrahedral junction at the bridge ground node, we plan to further suppress potential differences in the *Lo* leads of the detector, voltage sources, and $R_0$ leads from the QHARS.

Three resistors ($R_1$, $R_2$ and $R_0$) comprise the unknown resistance $R_X$, as seen in Fig. 1 (b). The 1.01 M$\Omega$ device is put in the place of $R_X$ (see the upper inset), with each of its two arrays, composed of 39 elements each and connected in series, meeting at a common node with a single element. The single element (valued at about 12.9 k$\Omega$) represents the *Gnd* terminal (or $R_0$, color coded green), whereas the two larger arms make up the *Hi* and *Lo* ($R_1$ and $R_2$, respectively) terminal connections.

In terms of the upper arm, which hosts the reference resistor $R_S$, calibration was necessary in order to accurately measure the QHARS device's transformed quantized value. As such, $R_S$ was calibrated against the NIST quantized Hall resistance (QHR) national standard with corresponding resistance bridges (and were valued at 1 M$\Omega$ and 10 M$\Omega$). This calibration history spans years and includes calibration data taken with both graphene- and GaAs-based QHRs. Both resistors have drift rates of 0.7 ($\mu\Omega/\Omega$)/yr or less and temperature coefficients of 0.2 ($\mu\Omega/\Omega$)/°C or less. The drift rates were determined by linear regression of historical data and the resistors' temperature was controlled to within ± 0.01 °C of 23 °C.

Before these high resistance experiments can commence, one preferred step for all QHARS devices is to assess their transport properties. This preliminary step helps optimize use of the more complicated DSB setup and measurements. These more basic quantum Hall transport measurements were performed with a Cryomagnetics C-Mag $^4$He cryostat (see Acknowledgements). All devices were mounted onto a transistor outline (TO-8) package, and all corresponding magnetoresistance data were collected between magnetic field values of 0 T and 8 T and at 2 K.

### IV. Measurement Results and Discussion

A pair of basic transport measurements is shown in Fig. 2 (a). The two values were measured to be close enough to their nominal values (corresponding to 78 and 40 elements for the black and red curves, respectively) that precision measurements were then warranted. The magnetoresistances in Fig. 2 (a) were collected with an HP 3458 digital voltmeter (see Acknowledgments). Though this technique allows one to collect data with higher magnetic field resolution, it potentially introduces small errors due to equipment impedance.

To perform a more precise measurement of the 1 MΩ QHARS, a CCC was used to make a two-terminal measurement. A nominal turns-ratio of 780:10, with a primary current of 0.775 μA, was applied to a 12.906 kΩ standard resistor using a cycle time of 60 s. A nominal current of 10 nA was applied to the 1 MΩ QHARS which was at a temperature of about 2.5 K while the magnetic field was swept from ± 2.8 T to ± 9 T. Thirty CCC measurements were made at each magnetic field, with the last sixteen measurements averaged for each field. Deviations from the nominal quantized value for the *Hi-Lo* 1 MΩ array are plotted in Fig. 2 (b) and show a comfortable approach to quantization just under 4 T.

(a)
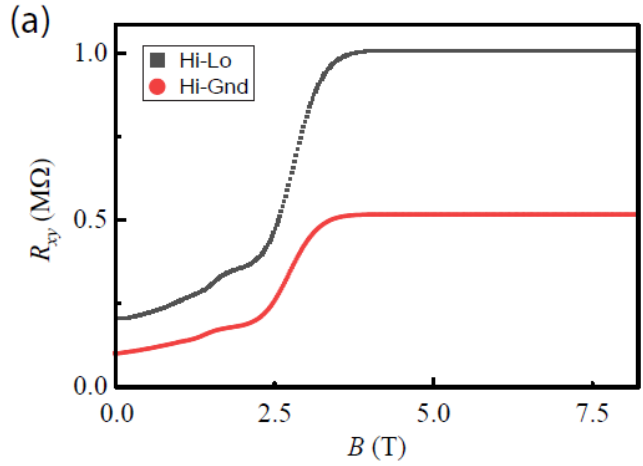

(b)
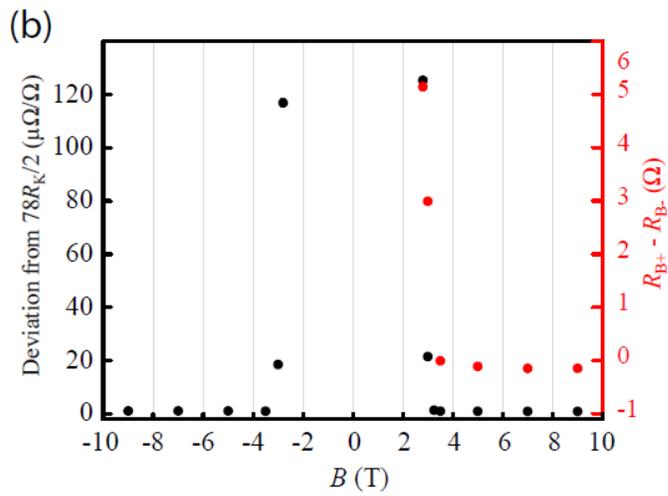

(c)
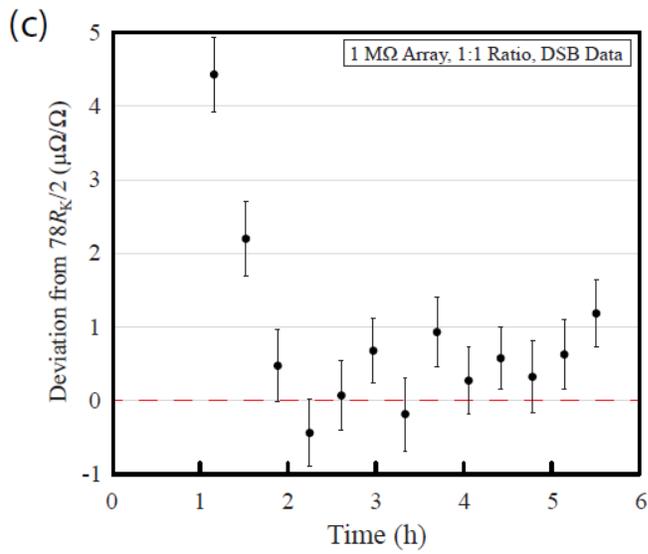

Fig. 2. (a) Basic transport measurements collected with an HP 3458 digital voltmeter. On the plateau (whose onset is about 4 T), the two values were measured to be close to their nominal values to within the measurement capability, as seen by the *Hi-Lo* (78 elements, black curve) and *Hi-Gnd* (40 elements, red curve). (b) CCC measurements taken at various magnetic field values verified the quantization of the QHARS device. The difference between the plateaus of opposite magnetic polarities is plotted in red with a right-side vertical axis. Error bars for data are smaller than the points. (c) Time-dependent DSB measurements are shown and use a 1:1 ratio against the reference resistor. Error bars represent the combined standard uncertainty ($k = 1$).

The difference between the plateaus obtained at opposite magnetic polarities is plotted in red with a right-side vertical axis. It is important to note the difference observed and its history of having been discussed in other work [33]. The device may not have been fully quantized, or a connection problem may have existed. When the field is reversed, the current goes through a different set of contacts, and the 2-terminal CCC measurement is more susceptible to these possible differences in contact resistance.

After the precision measurements demonstrated the metrological viability of the QHARS subarrays, the device was implemented into the DSB setup as shown in Fig. 1 (b). The balancing results of this method, which reflect a 1:1 ratio against the reference resistor, are shown in Fig. 2 (c). This time-dependent measurement validates the stability of this technique for these high resistances, after some time for settling, with some deviations only being off by a few parts in $10^7$. The final test is to prove the concept that this 1.01 MΩ device, while in the Y-network configuration, can exhibit the mathematically transformed value of resistance.

For the final test of the Y-Δ transformation, the test voltage applied to array was limited to a maximum of 10 V in order to protect cryostat wiring. Based on the transformation calculation, a value near 20.6 MΩ was predicted to be exhibited by the QHARS device when in a proper configuration. This value prompted the use of a 10 MΩ resistor, meaning an applied voltage ratio of about 2.06:1 could be applied. The time-dependent results of this measurement are shown in Fig. 3 (minus the first two points that fall off-scale but suggest a settling time of about 1.5 h). Since the nominal value for the plots and calculations was defined to be 20.6 MΩ exactly, a correction of 1812.6 μΩ/Ω should be applied to the vertical axis when calculating an absolute deviation from the quantum mechanical value (*i.e.*, the exact QHARS value near 20.6 MΩ). This exact value is demarcated as a dashed red line, and the blue dashed line is the average DSB result that excludes the initial settling measurement. The shaded blue indicates the standard deviation of the mean of those measurements.

The 5 μΩ/Ω offset from the theoretical value in the proof of principle experimental results may be attributed to the rudimentary DSB to QHARS connections where voltage differences at the connections to the QHARS are critical. Improvement to the bridge ground connection by using tetrahedral junctions and additional shielding would reduce lead resistance, thermals, and voltage drops for the measurement of the Y-Δ transformed QHARS. The current to $R_0$ flows mostly in one lead from the QHARS device (specific to the magnetic field orientation) and should be connected as close to the *Lo* terminal of $V_2$ as possible. A new ground junction box has been designed to improve the DSB to QHARS connection at the *Gnd* terminal using several tetrahedral junctions, which have been used in resistance standards to reduce cross junction resistance to 2 x $10^{-7}$ Ω or less [32]. Additionally, the 1 MΩ /100 kΩ and the 10 MΩ / 1 MΩ ratios were measured for the standard resistors (calibrated with CCC) on the DSB to investigate the 5 µΩ/Ω offset. Since this test did not reproduce the offset, one cannot correlate it to the worst case (maximum) 0.1 µΩ/Ω internal resistance of $V_2$.

One major source of uncertainty at 20.6 MΩ is the instability of the resistance ratio over long times. It is difficult to clearly assign this instability, despite its linearity, to the bridge connections and grounding circuit, but it is at least suggestive given the similar drift that occurs slowly (that is, over the course of hours). It is possible that thermal voltages fluctuate with similar time scales. In the event that one can optimistically treat this linearity as a systematic and predictable error, despite not knowing its origin with full certainty, it may be possible to mitigate or, in the less optimal case, use it to correct measured data.

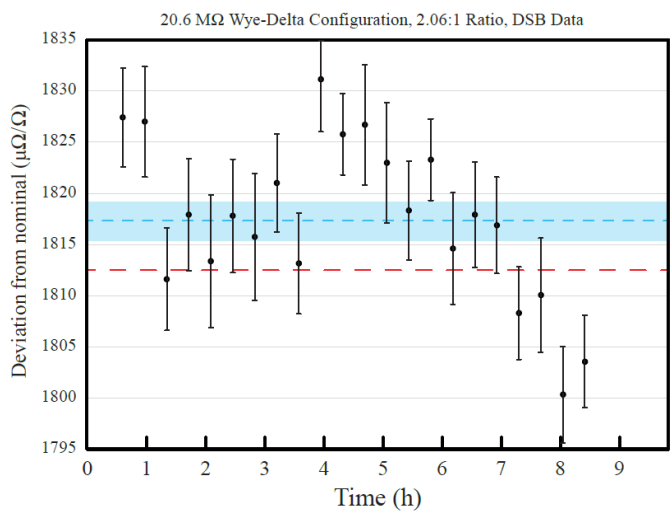

Fig. 3. Based on the Y-Δ transformation calculation, the 1.01 MΩ QHARS device was predicted to exhibit a resistance near 20.6 MΩ when in a proper configuration. A 10 MΩ resistor was used as $R_S$ in order to apply a voltage ratio of about 2.06:1. The time-dependent results shown a necessary settling time before relative stability. The theoretical QHARS value near 20.6 MΩ is shown as a dashed red line, and the blue dashed line is the average of a set of DSB

measurements. The shaded blue indicates the standard deviation of the mean of those measurements. Error bars represent the combined standard uncertainty ($k = 1$).

For now, the best case of calculating any deviation from the nominal value will inherently be dependent on the time of the measurement. The results in Fig. 3 were used to calculate the standard deviation of the mean, which itself has the future potential to be reduced to uncertainties of about 1 µΩ/Ω (or better should the drift issue be fully understood). Considering that typical 10 MΩ and 100 MΩ calibration measurements yield standard uncertainties of 1.3 µΩ/Ω and 1.6 µΩ/Ω, these results highlight the proof of concept that the Y-Δ transformation may be used to drastically reduce the calibration chain as well as provide a means to generate new quantum standards with many accessible high resistances depending on the measurement configuration.

One point of improvement for future devices and metrological studies would be to focus on maintaining the highest material quality for the relatively smaller arm (*Gnd*). Further, the use of a connector like those used in Hamon networks is critical to reduce errors for the $R_0$ resistor, which is more comparable to the resistance of the leads and connections than the other two arms. The use of the equalizing (4-way) connector would provide a better grounding, as defined by that of the bridge. In this case, any error stemming from the single quantum Hall element would have more drastic error ramifications due to the resistance's placement in the denominator of the Y-Δ transformation.

When inspecting two resistance networks containing $n$ terminals, like the one shown in Fig. 1 (a) and Fig. 4, one can derive a mathematical relationship between a star network (that is, all arms meeting at a central node like the left side of Fig. 1 (a) and Fig. 4) and its equivalent mesh network (where *n* is the same, but there exists one fewer node like the right side of Fig. 1 (a) and Fig. 4) [26], [34] – [35]:

$$R_{ik} = R_i R_k \sum_{\alpha=1}^{n} \frac{1}{R_\alpha}$$

(2)

In Eq. 2, the indices go as high as *n* and $i \neq k$. To double check the validity of this generalization, one can derive Eq. 1 in a straightforward manner (using *i, j,* and *k* as the indices). When a star has more than three terminals, it may also follow, for all indices, that:

$$\frac{R_{jk}}{R_{ik}} = \frac{R_{jl}}{R_{il}}$$

(3)

This condition must only be met in the event one wishes to transform a mesh to an equivalent star, and such a transformation is not always guaranteed. If one applies Eq. 2 to Fig. 4 (a) (*n* = 4), then:

$$R_{ij} = R_i + R_j + \frac{R_i R_j}{R_k} + \frac{R_i R_j}{R_l}$$

(4)

Just as in this main experimental work, by adding a similarly small resistor in parallel for two of the star arms, the equivalent resistance from the Y-Δ configuration nearly doubles. This favorable multiplicative attribute enables one to build quantum electrical standards with resistances as high as 10 GΩ, as shown below in Table II, especially since 13 parallel Hall bar elements have been demonstrated before [7], as have QHARS devices with 236 elements [14].

A star-mesh transformation for a 10 GΩ quantum standard is illustrated in Fig. 4 (d). Additional illustrations of potential circuit diagrams using a different star type are provided in the Appendix. Lastly, potential configurations are provided in Table II to show how adaptable this method is for high resistance traceability. One example of the potential of element reduction comes from the 10 GΩ case, where $7.75 \times 10^5$ elements in series are reduced, by means of the star-mesh transformation, to merely 502 elements.

TABLE II
APPLICABLE STAR-MESH TRANSFORMATIONS FOR FUTURE QHARS DEVICES

| $R_i$ (elements) | $R_j$ (elements) | $R_k - R_n$ (single-elements in parallel) | Total (elements) | R (MΩ) |
|---|---|---|---|---|
| 44 | 43 | 2 | 89 | 49.9607 |
| 50 | 50 | 3 | 103 | 98.0887 |
| 44 | 44 | 4 | 92 | 101.083 |
| 49 | 47 | 5 | 101 | 149.856 |
| 44 | 43 | 6 | 93 | 99.9407 |
| 139 | 139 | 4 | 282 | 1 001.05 |
| 188 | 187 | 11 | 386 | 4 995.95 |
| 244 | 244 | 13 | 501 | 9 995.44 |
| 245 | 244 | 13 | 502 | 10 036.4 |

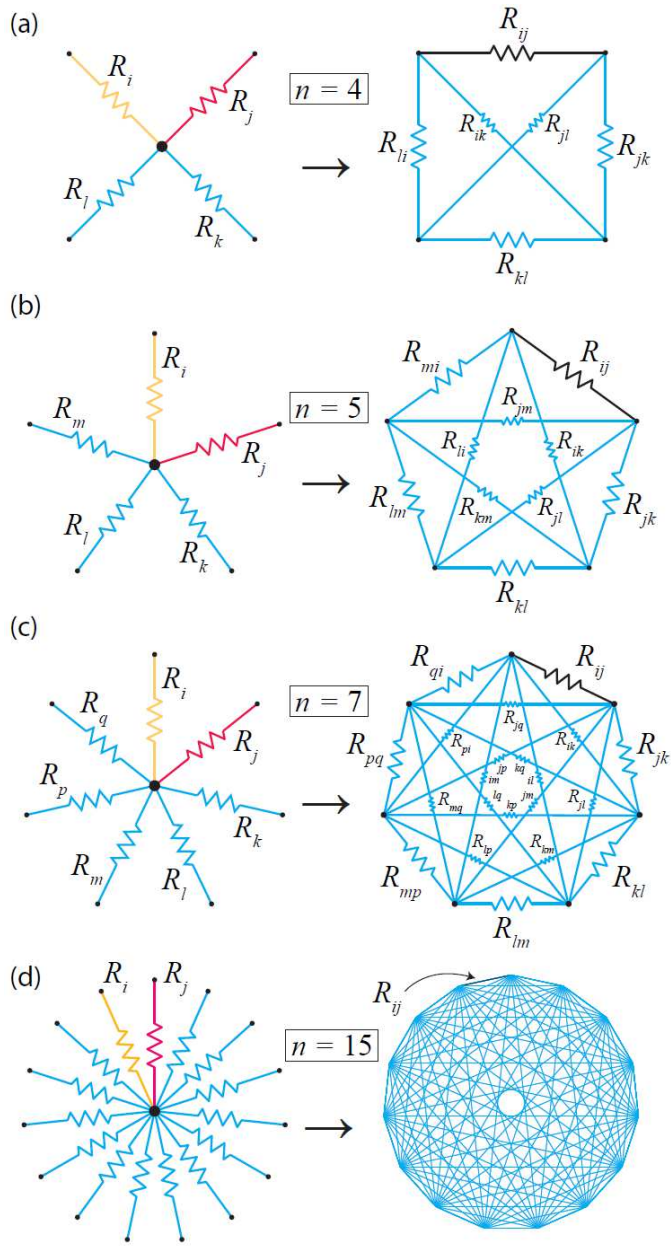

Fig. 4. (a) A 4-terminal star being transformed to a square mesh with four nodes. (b) A 5-terminal star transformed into a pentagonal mesh. (c) A 7-terminal star transformed into a heptagonal mesh. (d) A star-mesh transformation for a 10 GΩ quantum electrical standard is illustrated, representing a possible configuration that is within fabrication capacities, as seen in other work. The exact details of the configuration are provided in Table II. Two of the 15

20resistors in the star network are series arrays of several hundred resistors and the other 13 resistors are single Hall bar elements in parallel. For all subfigures, the uniform cyan color indicates the same potential, like ground as in this study, and applies to all mesh resistors except the high quantized resistance of interest.

## V. CONCLUSIONS

A 1.01 MΩ graphene-based QHARS device has been fabricated and shown to operate as an equivalent quantized resistor valued at about 20.6 MΩ by means of using a Y-Δ transformation and corresponding measurement configuration. This potent combination of using graphene-based technology with a mathematical transformation provides a way to extend QHR standards three decades beyond the 1 MΩ range. Additional values that may be attainable reach as high as 10 GΩ, rendering the Y-Δ transformation an incredibly efficient tool for reducing the required number of quantum Hall elements in series, at least, for resistances higher than 1 MΩ. The results presented herein are a proof of the concept that this type of circuit is beneficial to future resistance metrology applications.

## Appendix

Various optical images of example devices are shown in Fig. A1 and Fig. A2, with a light blue and orange region in the latter indicating the two example Hall elements whose Raman map results are shown in Fig. A3 (a) and (b), respectively.

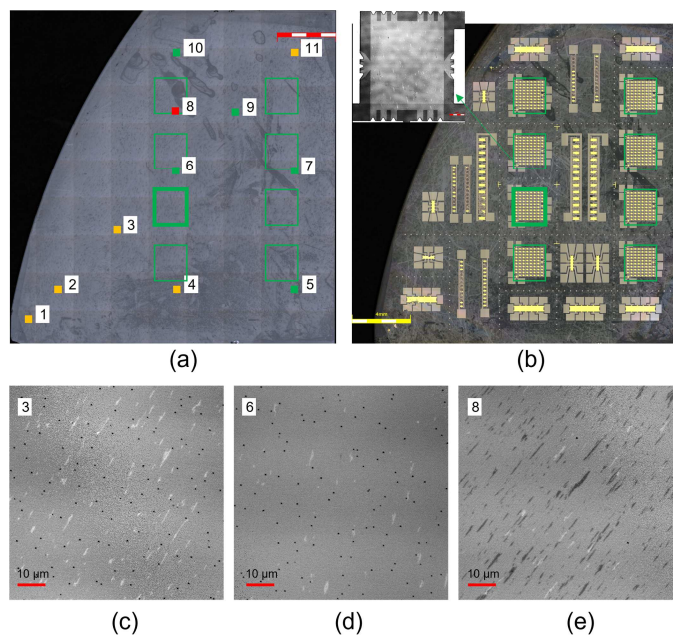

Fig. A1. (a) An optical image is shown of the full SiC chip with EG before fabrication. The red scale bar at the top right corner represents 4 mm. (b) The same region is shown post-fabrication. The yellow scale bar at the left bottom corner represents 4 mm. (c) Confocal images are shown for: site 3 (orange spot), showing full monolayer EG coverage with some multilayers, (d) site 6 (green square), showing minimal multilayer graphene, and (e) site 8 (red spot), showing incomplete monolayer EG with some existing buffer layer.

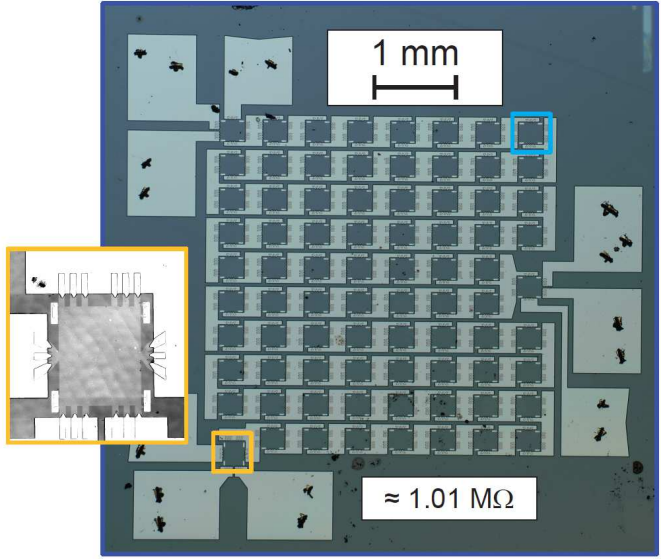

Fig. A2. An optical image of an example device. The inset for the orange region is also provided to demonstrate both the multi-series connections and the clarity with which CLSM successfully identifies EG monolayers.

For robust statistics on the quality of the EG films, rectangular Raman maps were collected with step sizes of 1 μm in a 25 × 25 raster-style grid and repeated on the two outermost corner elements of an example array device. Each spectrum exhibited a clear 2D (G') peak, which was subsequently fit with a Lorentzian profile to extract a peak position and full-width-at-half-maximum (FWHM). These quantities were used as the primary metric for comparing EG quality across the devices. It should be noted that the D and G peaks were not selected for determining homogeneity because their spectral neighborhood is strongly dominated by optical responses from the SiC substrate [24]. The resulting scatterplot for one of the elements is shown in Fig. A3 (a). For the other element (Fig. A3 (b)), a spatial map is presented with values of the FWHM to give a better visualization of the variation in optical response within that region. Three gray spots on this map represent minor bilayer growths that generally do not affect the quality of electrical measurements. Overall, these data confirm the length scales on which EG can be grown with excellent quality.

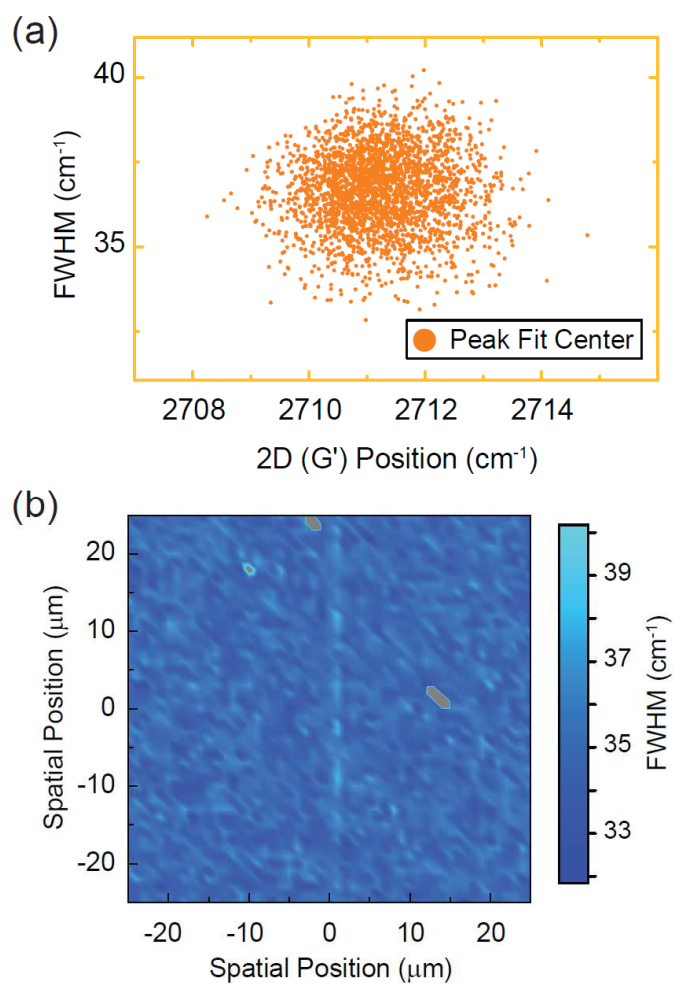

Fig. A3. (a) A scatterplot of the 2D (G') Raman mode of graphene places peak position ($k$-space) against full-width at half-maximum (FWHM), helping to verify homogeneity. Each measured peak was fitted with a Lorentzian profile. (b) A spatial map of the FWHM for the light blue box

in Fig. A2 demonstrates the uniformity of the EG film. Three gray spots indicate minor bilayer growths that generally do not affect the quality of electrical measurements.

Other circuit designs could implement additional parallel branches with contact pads that are bonded together during fabrication. One such example is seen in Fig. A4. In this case, the device is presumed to measure 0.86 M$\Omega$ across the sum of $R_i$ and $R_j$, but after performing a star-mesh transformation, is calculated to provide a value of about 27.3 M$\Omega$. This example device is composed of 32 elements for each of two larger branches and connects in series with a common node that also meets with two distinct and additional branches, each containing a single element. The single element represents the *Gnd* (or $R_0$, color coded cyan), whereas the two larger arms make up the *Lo* and *Hi* ($R_1$ and $R_2$, respectively) terminal connections. Though these values were arbitrarily chosen, the exemplify a benefit in using additional grounded branches as a means to reduce the number of required devices to achieve large transformed quantized resistances.

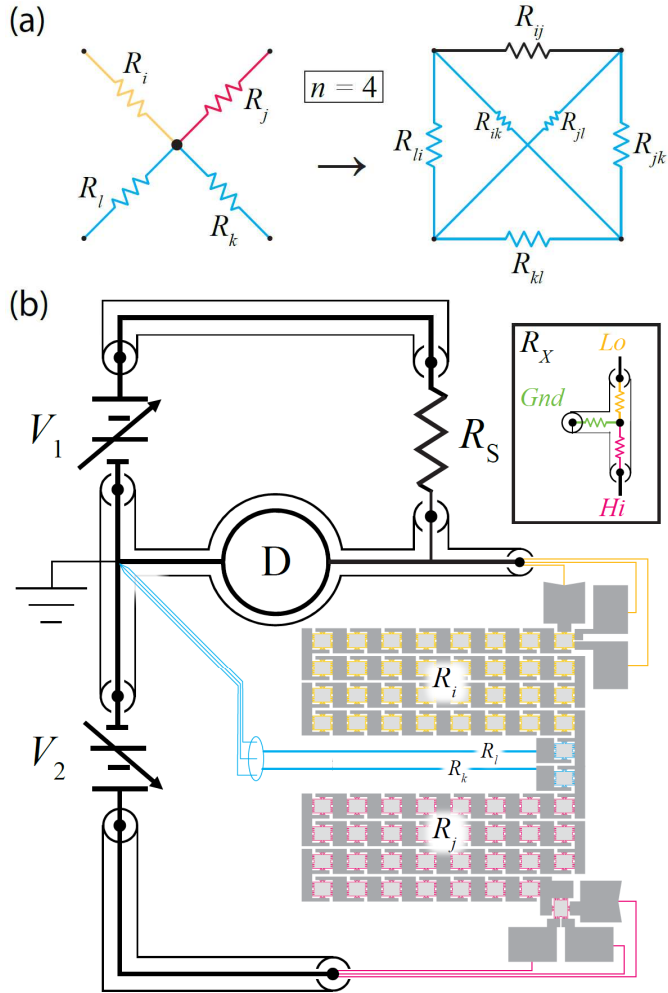

Fig. A4. (a) A 4-terminal star being transformed to a square mesh with four nodes. (b) Simplified DSB diagram for a different QHARS. The device, calculated to provide a transformed value of 27.3 MΩ (and presumed to measure 0.86 MΩ across the sum of $R_i$ and $R_j$) and drawn for sake of example, is intended to substitute $R_X$ (see upper inset), with each of its two arrays, composed of 32 elements each and connected in series, meeting at a common node with two distinct and additional branches, each containing a single element. The single element represents the *Gnd* (or $R_0$, color coded cyan), whereas the two larger arms make up the *Lo* and *Hi* ($R_1$ and $R_2$, respectively) terminal connections.

26## ACKNOWLEDGMENTS AND NOTES

The authors would like to thank L. Chao, F. Fei, and E. C. Benck for their assistance during the internal review process at NIST. The work of S. Mhatre and C.-C. Yeh at NIST was made possible by arrangement with Prof. C.-T. Liang of National Taiwan University.

Certain commercial equipment, instruments, or materials are identified in this paper to foster understanding. Such identification does not imply recommendation or endorsement by the National Institute of Standards and Technology or the United States Government, nor does it imply that the materials or equipment identified are necessarily the best available for the purpose.
REFERENCES

[1]    T. J. Janssen, J. M. Williams, N. E. Fletcher, R. Goebel, A. Tzalenchuk, R. Yakimova, S. Lara-Avila, S. Kubatkin, V. I. Fal'ko, "Precision comparison of the quantum Hall effect in graphene and gallium arsenide," Metrologia. 49, 294, 2012.

[2]    J. Kučera, P. Svoboda, K. Pierz, "AC and DC quantum Hall measurements in GaAs-based devices at temperatures up to 4.2 K," IEEE Trans. Instrum. Meas., 68, 2106-12, 2018.

[3]    A. F. Rigosi and R. E. Elmquist, "The Quantum Hall Effect in the Era of the New SI," Semicond. Sci. Technol., 34, 093004, 2019.

[4]    A. Tzalenchuk, S. Lara-Avila, A. Kalaboukhov, S. Paolillo, M. Syväjärvi, R. Yakimova, O. Kazakova, T. J. Janssen, V. Fal'ko, S. Kubatkin, "Towards a quantum resistance standard based on epitaxial graphene," Nat. Nanotechnol., 5, 186, 2010.